\documentclass[12pt]{article}

\usepackage{amssymb}
\usepackage{amsmath}
\usepackage{latexsym}

\catcode`@=11
\renewcommand{\section}{\@startsection{section}{1}{0pt}{\medskipamount}
{\medskipamount}{\large\bf}}
\numberwithin{equation}{section}
\catcode`@=12

\def\beq{\begin{eqnarray}}    
\def\eeq{\end{eqnarray}}      

\def\tr{\,\mbox{tr}\,}                  

\def\pa{\partial}                       

\def\={\ =\ }

\begin{document}

\begin{center}
{\Large\bf Representation of a gauge field via intrinsic "BRST" operator }

\vspace{18mm}

{\Large Igor A. Batalin$^{(a,b)}\footnote{E-mail:
batalin@lpi.ru}$\;,
Peter M. Lavrov$^{(b, c)}\footnote{E-mail:
lavrov@tspu.edu.ru}$\;
}

\vspace{8mm}

\noindent ${{}^{(a)}}$
{\em P.N. Lebedev Physical Institute,\\
Leninsky Prospect \ 53, 119 991 Moscow, Russia}

\noindent  ${{}^{(b)}}
${\em
Tomsk State Pedagogical University,\\
Kievskaya St.\ 60, 634061 Tomsk, Russia}

\noindent  ${{}^{(c)}}
${\em
National Research Tomsk State  University,\\
Lenin Av.\ 36, 634050 Tomsk, Russia}

\vspace{20mm}

\begin{abstract}
\noindent We show that there exists a representation of a matrix
valued gauge field via intrinsic "BRST" operator assigned to matrix
valued generators of a gauge algebra.  In  this  way, we reproduce
the standard formulation of the ordinary Yang - Mills theory.  In
the case of a  generating quasigroup/groupoid, we give a natural
counterpart to the Yang - Mills action. The latter counterpart does
also apply as to the most general case of an involution for
matrix-valued gauge generators.
\end{abstract}

\end{center}

\vfill

\\

\newpage

\section{Introduction and summary}

All modern models describing the fundamental forces in the Nature are based
on the concept of gauge fields \cite{YM,DeW1,DeW,FP,FT,T,S,DeWitt}.
It  is  a  well-known  fact that the BRST symmetry \cite{brs1,t} is the most powerful
method  to represent  the invariance properties of a  gauge field system \cite{KO,BV}.
Usually, in simple examples,  in  Hamiltonian formalism,  gauge generators
have the form of secondary constraints similar to the "Gauss law"
represented as covariant divergence of canonical momenta. These generators
are in involution that represents a gauge algebra on the phase
space of the system \cite{BF1,BF2}. By introducing ghost canonical pairs,  one is able to
define the respective nilpotent BRST operator containing the first class
constraints in its lowest terms, linear in ghost coordinates.  In the
respective Lagrangian formalism, the gauge generators are represented in
terms of Lagrangian field variables, as  the coefficients linear in the
original antifields, entering the minimal master action. In this way,
usually, space-like components of relativistic fields are identified with Hamiltonian
coordinates , while time - like components are identified with Lagrange
multipliers  to secondary first-class constraints. In the simplest example,
the Yang - Mills theory,  Lagrangian matrix-valued gauge field is a linear
combination of matrix valued generators of adjoint representation of a
generating Lie group.  Thus, if one has defined the respective intrinsic
"BRST" operator assigned to matrix valued generators of adjoint
representation, one can define the matrix valued gauge field as a commutator
of intrinsic "BRST" operator with an auxilliary "gauge" Fermion linear in
the adjoint component of the Yang - Mills field . Thus, one has arrived at
the intrinsic  "BRST" representation to the matrix  valued  gauge field.

In the present article, we study in detail the approach based on the
intrinsic "BRST" representation. In the case of a Lie group we have
shown that the new approach  does reproduce exactly the standard
formulation of the  Yang  - Mills  theory. Then, we consider the
case of a quasigroup/groupoid \cite{Sab,M,B,BVjmp,CFer,Ber,MSt,GSt},
where the structure coefficients of the  intrinsic algebra of matrix
valued generators,  are matrix valued themselves. In that case, we
have found a natural counterpart to the Yang-Mills action.  Finally,
we consider the most general  case of  being the matrix valued
generators in the general involution among themselves.

\section{Outline of the construction}

Let $A_{\mu}(x)$ be a Boson  $N\times N$ matrix valued vector field
as defined by the formula
\beq
\label{j1}
A_{\mu} (x)  =  [
{\cal A}_{\mu} (x), Q  ],\quad [  A_{\mu} (x), Q  ]  = 0,
\quad\varepsilon ( A_{\mu} )  =  0, \quad
{\rm gh} ( A_{\mu} )  =  0, 
\eeq
where $Q$ is a nilpotent Fermion operator,
\beq
\label{j2}
Q^{2}  =  \frac{1}{2} [ Q , Q ]  =  0, \quad  \varepsilon( Q )  =  1,   \quad
{\rm gh}( Q )  =  1, 
\eeq
and Fermion  vector field ${\cal A}_{\mu} (x)$  has ghost number $-1$,
\beq
\label{j3}
\varepsilon ( {\cal A}_{\mu} ) = 1,  \quad
{\rm gh} ( {\cal A}_{\mu} )  =  -1. 
\eeq
In more detail, $Q$ is an $N \times N$ matrix valued operator and, at
the same time does depend on $m$ Fermion  canonical  pairs  of
ghosts $( C^{a}, \mathcal{P}_{a} )$, $a = 1, ..., m$,  $\varepsilon(
C^{a} )  =  \varepsilon ( \mathcal{P}_{a} )  = 1$, ${\rm gh}( C^{a}
)  =  1$, ${\rm gh}( \mathcal{P}_{a} ) =  -1$, \beq \label{j4} [
C^{a}, C^{b} ]  =  0, \quad [ C^{a}, \mathcal{P}_{b} ]  =
\delta^{a}_{b},  \quad [ \mathcal{P}_{a}, \mathcal{P}_{b} ]  =  0.   
\eeq
We assume these operators to be realized as $n \times n$ matrices, so that in
fact the $Q$  is defined on tensor product of the original matrix
arguments and the ones of ghosts in (\ref{j4}). The same status
we do assume as to the ${\cal A}_{\mu}$
in (\ref{j1}), (\ref{j3}). The assumption of a matrix realization of ghosts,
together with the relations (\ref{j4}),  allows one to get simple expressions
for traces of homogeneous
$C \mathcal{P}$ normal ordered  monomials with ghost number zero. In
what follows, we will  use the two simple examples,
\beq
\label{j5}
\tr( C^{a} \mathcal{P}_{b} )  =  \frac{n}{2} \delta^{a}_{b},   
\eeq
\beq
\label{j6}
\tr( C^{a} C^{b} \mathcal{P}_{c} \mathcal{P}_{d} )  =  \frac{n}{4} \left(
\delta^{a}_{d} \delta^{b}_{c}  - \delta^{a}_{c} \delta^{b}_{d} \right).    
\eeq
These two formulae do follow from the general representation for ghost
canonical pairs in terms of two conjugate sets of $n\times n$ gamma matrices,
\beq
\label{j6*}
2 C^{a}  =  \gamma_{+}^{a}  +  \gamma _{-}^{a},  \quad
2 \mathcal{P}_{a}   =  (\gamma_{+}^{b}  -  \gamma_{-}^{b} )g_{ba},        
\eeq
where $g_{ab}  =  g _{ba}$ is a constant invertible metric,  $g^{ab}  =  g
^{ba}$  is its inverse, and  the $\gamma$ matrices do commute as
\beq
\label{j7*}
\gamma_{\pm}^{a} \;\gamma_{\pm}^{b}  +  ( a \leftrightarrow b)&=&
(\pm) 2 g^{ab}\;\! {\bf 1}, \\
\label{j8*}
\gamma_{\pm}^{a} \;\gamma_{\mp}^{b}  +  \gamma_{\mp}^{b}\; \gamma_{\pm}^{a}
&=& 0.    
\eeq
It follows from (\ref{j6*}) that (\ref{j5}), (\ref{j6}) do
generalize to
\beq
\label{j1**} \tr ( X ( C,  \mathcal{ P } ) )  =
\left( X \left( \frac{\pa}{ \pa J},  \frac{\pa}{ \pa K} \right)
\exp\left\{\frac{1}{2} K^{a} J_{a}  \right\} n  \right) \Big|_{ J = 0,  K = 0 } \;,     
\eeq
\beq \label{j2**} \varepsilon( J )  =  1 ,\quad   {\rm gh}( J )
=  - 1, \quad \varepsilon( K )  =  1,\quad {\rm gh}( K )  =  1,
\quad
\varepsilon ( X )  =  0, \quad   {\rm gh} ( X )   =   0.   
\eeq

By inserting the doublet (exact) form, the first in (\ref{j1}), into the
curvature form
\beq
\label{j10*}
G_{\mu\nu} = \pa_{\mu} A_{\nu} - \pa_{\nu} A_{\mu} +
[ A_{\mu}, A_{\nu} ],    
\eeq
we have
\beq
\label{j11*}
G_{\mu\nu} = [ Q, \mathcal{G}_{\mu\nu} ],      
\eeq
\beq
\label{j12*}
\mathcal{G}_{\mu\nu} = \pa_{\mu} \mathcal{A}_{\nu} -
\pa_{\nu}\mathcal{A}_{\mu}
+ ( \mathcal{A}_{\mu}, \mathcal{A}_{\nu} )_{Q},       
\eeq
where in (\ref{j12*}) the quantum antibracket $(X, Y)_{Q}$
is defined for any two operators $X, Y$,  as \cite{BM}
\beq
\label{j13*}
2 ( X, Y )_{Q} = [ X, [ Q, Y ] ] - [ Y, [ Q, X ] (-1)^{ (\varepsilon_{X}
+1) (\varepsilon_{Y} + 1 ) }.     
\eeq
 When deriving (\ref{j12*}), we have used the general property
\beq
\label{j14*}
 [ [ Q, X ], [ Q, Y ] ] = [ Q, ( X, Y )_{Q} ]=
 ( [Q, X ], Y )_{Q}  -  ( X, [Q, Y] )_{Q} (-1)^{ \varepsilon_{ X } }.   
\eeq
Notice that the quantum antibrackets do satisfy the Jacobi
identity modulo a doublet (exact) form \beq \nonumber &&( X, (Y,
Z)_{Q} )_{Q} (-1)^{ (\varepsilon_{X} + 1) (\varepsilon_{Z} + 1) }
+ {\rm cyclic \;perm.} ( X, Y, Z ) =\\
\label{j15*} &&= \frac{1}{2} [ (X, Y, Z )_{Q} (-1)^{
(\varepsilon_{X} + 1) (\varepsilon_{Z}
+1) }, Q ],    
\eeq
where $( X, Y, Z )_{Q}$  is the so-called quantum 3-antibracket,
\beq
\nonumber
&&3 ( X, Y, Z )_{Q}  = - (-1)^{ (\varepsilon_{X} + 1) (\varepsilon_{Z} + 1) }
(  [ X, ( Y, Z )_{Q} ] (-1)^{ [ \varepsilon_{X} (\varepsilon_{Z} + 1)  +
\varepsilon_{Y} ] }  +\\
\label{j16*}
 &&\qquad\qquad\qquad+{\rm cyclic \; perm}. ( X, Y, Z )  ),     
\eeq
and so on \cite{BM1} (see also \cite{K-S,Vo,CSch}).
The modified Leibnitz rule for quantum antibracket reads
\beq
\nonumber
&&( XY, Z )_{Q}  -  X ( Y, Z )_{Q}  - ( X, Z )_{Q} Y (-1)^{ \varepsilon_{Y} (
\varepsilon_{Z} + 1 ) } = \\
\label{j217}
&&=\frac{1}{2} ( [ X, Z] [ Y, Q ] (-1)^{ \varepsilon_{Z} ( \varepsilon_{Y} + 1 ) }
+  [ X, Q ] [ Y, Z ] (-1)^{ \varepsilon_{Y} } ).       
\eeq

In terms of the curvature (\ref{j11*}), the General  "Yang-Mills" Lagrangian reads
\beq
\label{j16*}
 \mathcal{L} = - \frac{1}{2} \tr ( G_{\mu\nu} G_{\mu\nu} ) =
- \frac{1}{2} \tr ( [Q,\mathcal{G}_{\mu\nu} ] [ Q, \mathcal{G}_{\mu\nu} ] ),       
\eeq

Let us consider infinitesimal gauge transformations with an operator
valued
Fermion "parameter" $\Xi$,  $
\varepsilon(\Xi)= 1$,  ${\rm gh}(\Xi) = -1$,  $\Xi \;\rightarrow \;0$,
\beq
\label{j17*}
\delta A_{\mu} = - ( \pa_{\mu} [ Q, \Xi ] + [ A_{\mu}, [Q, \Xi ] ] ). 
\eeq
It follows from the first in (\ref{j1}), and (\ref{j14*}), that the respective
 variation in $\mathcal{A}_\mu$ can be chosen in the form
\beq
\label{j18*}
\delta \mathcal{A}_{\mu} = - ( \pa_{\mu} \Xi +
( \mathcal{A}_{\mu}, \Xi )_{Q}).   
\eeq
Due to (\ref{j12*}), (\ref{j15*}), it follows that the respective variation in
$\mathcal{G}_{\mu\nu}$ can be chosen in the form
\beq
\label{j19*}
\delta \mathcal{G}_{\mu\nu} = -  ( \mathcal{G}_{\mu\nu}, \Xi )_{Q}. 
\eeq
Now, we have, as to the respective variation in (\ref{j16*})
\beq
\nonumber
&&\delta \mathcal{L} = -\tr ( [Q, \delta \mathcal{G}_{\mu\nu} ] [ Q,
\mathcal{G}_{\mu\nu} ] ) =
\tr ( [ [ Q, \mathcal{G}_{\mu\nu} ] , [Q, \Xi ] ] [ Q,
\mathcal{G}_{\mu\nu} ] ) = \\
\label{j20*}
&&= \tr ( [ Q, \mathcal{G}_{\mu\nu} ] [Q, \Xi ]
[Q, \mathcal{G}_{\mu\nu} ]
- [ Q, \Xi ] [Q, \mathcal{G}_{\mu\nu} ] [ Q, \mathcal{G}_{\mu\nu} ] )= 0.
\eeq
Here, in the second equality we have used (\ref{j14*}) backward, and we have
 moved the last commutator to the leftmost position in the second term in the
left-hand side of the last (fourth) equality. Thereby, we have
confirmed explicitly  that the Lagrangian  (\ref{j16*}) is gauge
invariant. Thus, we have constructed a family of gauge-invariant
classical theories of the type (\ref{j16*}), closely related to the
"general Yang-Mills theory".  Every of those classical theories can
certainly be considered as a starting point as to apply the
Hamiltonian BFV or Lagrangian BV quantization scheme,  although we
do not do that in the present article.

In what follows below through the article, we assume the operator $Q$ as
represented in $C \mathcal{P}$ normal form.  In that case, it follows in terms
of the quantum antibrackets, with no further assumptions,
\beq
\label{j2.21}
( T_{a}, T_{b} )_{Q}  =  0,   
\eeq
 \beq
\label{j2.22}
( \mathcal{P}_{a}, \mathcal{P}_{b} )_{Q}  +  C^{c} ( \mathcal{P}_{c},
\mathcal{P}_{a}, \mathcal{P}_{b} )_{Q}  =  U_{ab}^{c} \mathcal{P}_{c},
\eeq
\beq
\label{j2.23}
2 ( \mathcal{P}_{a}, T_{b} )_{Q}  =  [ T_{a}, T_{b} ],      
\eeq
 where we have denoted
\beq
\label{j2.24}
T_{a}  =  [ \mathcal{P}_{a}, Q  ],   \quad  U_{ab}^{c}  =  [ [ \mathcal{P}_{a}, [
\mathcal{P}_{b}, Q ] ], C^{c} ].        
\eeq
 and  the quantum 3-antibracket of the ghost momenta reads
\beq
\label{j2.25}
( \mathcal{P}_{a}, \mathcal{P}_{b}, \mathcal{P}_{c} )_{Q}   = [
\mathcal{P}_{a}, [ \mathcal{P}_{b}, [ \mathcal{P}_{c}, Q ] ] ] .     
\eeq
In turn, by commuting the $Q$ with (\ref{j2.22}), we get
\beq
\label{j2.26}
[ T_{a}, T_{b} ]  +   [ Q, C^{c} ( \mathcal{P}_{c}, \mathcal{P}_{a},
\mathcal{P}_{b} )_{Q} ]  =  U_{ab}^{c} T_{c}  +  [Q, U_{ab}^{c} ]
\mathcal{P}_{c}.      
\eeq
 If we assume that $A_{\mu}^a$ and $\Xi^a$ are $c$-numbers, and
\beq
\label{j2.27}
\mathcal{A}_{\mu} = A_{\mu}^{a} \mathcal{P}_{a},
\quad \Xi  =  \Xi^{a} \mathcal{P}_{a}\quad\;\;\; ( {\rm Mod}\;  [ Q , {\rm Anything}  ] )  
\eeq
 then we get, due  to the first in (\ref{j1}) and in (\ref{j2.24}),
\beq
\label{j2.28}
A_{\mu} = A_{\mu}^{a} T_{a},\quad  [ \Xi, Q ]  =  \Xi^{a} T_{a}.      
\eeq
 If , moreover, $U_{ab}^c$ are $c$-numbers, and the metric,
\beq
\label{j2.29}
\eta_{ab} = (Nn)^{-1} \tr ( T_{a} T_{b} ),     
\eeq
 is invertible, so that $\eta^{ab}$ is its inverse, then we have for the field
components
\beq
\label{j2.30}
A_{\mu}^{a} =  (Nn)^{-1} \tr ( A_{\mu} T_{b} ) \eta^{ba},       
\eeq
 and, therefore, their gauge transformation presents
\beq
\label{j2.31}
\delta A_{\mu}^{a} =  - \pa_{\mu} \Xi^{a}  -  A_{\mu}^{c} \Xi^{d} (Nn)^{-1}
\tr ( [ T_{c}, T_{d} ] T_{b} ) \eta^{ba}.      
\eeq

\section{Yang-Mills theory generated by a compact semisimple Lie group}

Let  $t_{a}, a  =  1, ... ,m $, be $N\times N$ matrix valued  Boson
generators of a semisimple Lie group,
\beq
\label{j7}
[ t_{a}, t_{b} ]  =  U_{ab}^{c}  t_{c},  \quad   \tr( t_{a} ) = 0,
\eeq
where $U_{ab}^{c}  =  - U_{ba}^{c}  =  const$ are structure constants
of the group.  They satisfy the relations
\beq
\label{j8}
U_{ab}^{c} U_{cd}^{e}  +  {\rm cyclic \;perm.} ( a, b, d )  =  0,\quad
U_{ab}^{b} =  0.      
\eeq
Due to the first in (\ref{j7}) and (\ref{j8}), the following operator
\beq
\label{j9}
Q  =  C^{a} t_{a}  + \frac{1}{2} C^{b} C^{a} U_{ab}^{c} \mathcal{P}_{c}, 
\eeq
does satisfy (\ref{j2}).  Vise versa, the  nilpotency condition (\ref{j2}) does imply the algebra
of the first (\ref{j7}) and (\ref{j8}).
Now, let us choose the operator ${\cal A}_{\mu} (x)$ in
the form
\beq
\label{j10}
{\cal A}_{\mu} (x) = A_{\mu}^{a}(x) \mathcal{P}_{a}.   
\eeq
It follows then from (\ref{j1})
\beq
\label{j11}
A_{\mu} (x)  =  A_{\mu}^{a} (x) T_{a},  \quad     T_{a}  =  [
\mathcal{P}_{a}, Q ]  =  t_{a}  +  C^{b} U_{ba}^{c} \mathcal{P}_{c},  
\eeq
\beq
\label{j11**}
\tr ( A_{\mu}(x) T_{b} )  =  A_{\mu}^{a}(x)  \tr ( T_{a} T_{b} ).   
\eeq
In turn, due to the first in (\ref{j7}) and (\ref{j8}), it follows for the second
in (\ref{j11})
\beq
\label{j12}
[ T_{a}, T_{b} ] = U_{ab}^{c} T_{c},  \quad    \tr( T_{a} ) = \frac{Nn}{2}
U_{ba}^{b} = 0.              
\eeq
Then, we have
\beq \nonumber
\tr( T_{a} T_{d} ) &= &n \tr(
t_{a} t_{d} ) + N \tr \left( C^{b}U_{ba}^{c}
\mathcal{P}_{c} C^{e} U_{ed}^{f} \mathcal{P}_{f}\right )  =\\
\label{j13} &=&
N n  \left[  N^{-1}  \tr ( t_{a} t_{d} )  +  \frac{1}{4} U_{ca}^{b} U_{bd}^{c}  \right],
\eeq
where (\ref{j4}), (\ref{j5}), (\ref{j6}) have been used. Thus, we have reproduced the
well-known Yang -Mills Lagrangian
\beq
\label{j14}
{\cal L} =- \frac{1}{2Nn}   \tr ( G_{\mu\nu}(x) G_{\mu\nu}(x) )  =
 - \frac{1}{2Nn}  G_{\mu\nu}^{a}(x) G_{\mu\nu}^{b}(x) \tr(
T_{a} T_{b} ),
\eeq
where the Yang - Mills curvature (stress tensor ) has the usual form
\beq
\label{j15}
G_{\mu\nu}(x) & = & \pa_{\mu} A_{\nu}(x)  -  \pa_{\nu} A_{\mu}(x)  +
[ A_{\mu}(x), A_{\nu}(x)]  =  G_{\mu\nu}^{a}(x) T_{a},  \\  
\label{j16}
G_{\mu\nu}^{a}(x) &=& \pa_{\mu} A_{\nu}^{a}(x) - \pa_{\nu} A_{\mu}^{a}(x)  +
A_{\mu}^{b}(x) A_{\nu}^{c}(x) U_{bc}^{a}.      
\eeq Ghost - extended generators,  similar to the second in
(\ref{j11}), have been first introduced in string theory
\cite{Hw,GShW}, and then generalized and studied systematically in
\cite{BTtmf,BB}, being called as "BRST-invariant constraints".

\section{The quasigroup/groupoid case}

Now, let us consider a more general situation of quasigroup/groupoid, where
the structure coefficient of the algebra are matrix-valued operators rather
then constants. In that case we have
\beq
\label{j17}
[  t_{a}, t_{b} ]  =  U_{ab}^{c} t^{c}, \quad  U_{ab}^{c} = - U_{ba}^{c},
\quad   \tr(U_{ab}^{c} t_{c} )  =  0,     
\eeq
\beq
\label{j18}
(  U_{ab}^{c} U_{cd}^{e}  -  [ t_{d}, U_{ab}^{e} ]  )  +  {\rm cyclic\;  perm.} ( a,
b, d )  =  0,       
\eeq
\beq
\label{j19}
(  [ U_{ab}^{c},  U_{de} ^{f} ]  -  ( c \leftrightarrow f ) )  +
{\rm cyclic \; perm.} ( a,b, d, e )  =  0,    
\eeq
where we have denoted
\beq
\label{j20}
X_{abcd}  +  {\rm cyclic \;perm.} ( a, b, c, d )  = 4! S_{abcd}^{hgfe}  X_{efgh}, 
\eeq
\beq
\label{j21}
4 ! S_{abcd}^{hgfe} = \pa_{a} \pa_{b} \pa_{c} \pa_{d} C^{h} C^{g} C^{f} C^{e},\quad
\pa_{a}  = \frac{\pa}{\pa C^{a}}.  
\eeq
Due to these relations  (\ref{j17}), (\ref{j18}), (\ref{j19}),
the operator (\ref{j9}) with the  matrix
valued structure operators $ U_{ab}^{c}$ does satisfy the nilpotency (\ref{j2}).
The quasigroup/groupoid is the most general case of generators, where  the
operator $Q $ (\ref{j9}) linear in ghost momenta  does satisfy the nilpotency (\ref{j2}).

If one defines a counterpart to the BRST-invariant generators, the second
in (\ref{j11}), with operator valued $U_{ab}^{c}$,
then the respective algebra
\beq
\label{j4.6}
[T_{a}, T_{b}] = U_{ab}^{c} T_{c} + [Q, U_{ab}^{c}]\mathcal{P}_{c},    
\eeq
\beq
\eta_{ab}  =  (Nn)^{-1}  \tr ( T_{a} T_{b} )  =
N^{-1} \tr \left[ \left( t_{a} + \frac{1}{2}
U_{ca}^{c} \right) \left( t_{b} + \frac{1}{2} U_{db}^{d} \right) +
\frac{1}{4} U_{da}^{c} U_{cb}^{d} \right],
\eeq
does involve the ghost momenta $\mathcal{P}_{a}$ to serve as
 new generators with their own semi-Abelian subalgebra,
\beq
\label{j4.7}
[\mathcal{P}_{a},\mathcal{P}_{b}] = 0,  \quad [\mathcal{P}_{a},T_{b}] =
U_{ab}^{c}\mathcal{P}_{c}=({\cal P}_a, {\cal P}_b)_{Q}.     
\eeq
Notice that if one commutes the $Q$ with (\ref{j4.6}), one gets no further
consequences. So, only the doubled set of generators,
\beq
\label{j4.8}
\mathcal{T}_{A} = \{ T_{a}; \mathcal{P}_{a} \},    
\eeq
does have a closed algebra. However the general formulation of
Section 2 appears capable to operate efficiently even in such a
complicated situation, as a part of the most general case to be considered below.
Here, we restrict ourselves by rewriting the general Lagrangian (\ref{j16*})
in the form more explicit in respect
of being the operator $Q$  (\ref{j9}) only linear in ghost momenta $\mathcal{ P }_{ a }$,
even when generalized as
for the matrix valued structure operators $U_{ ab }^{ c }$ specific
to the quasigroup /grouppoid case,
\beq
\label{j4.10}
\mathcal{ L }  =:  -  \frac{ 1 }{ 2 } \tr ( G_{ \mu\nu }^{ A } \mathcal{ T }_{ A}\;\!
G_{ \mu\nu}^{ B }  \mathcal{ T }_{ B } ),
\eeq
where, in the right-hand side, every capital index of the $"A"$ type
is split into two small indices of the $"a"$ type of
a half-dimension, $ A  =  \{ a ; a \}$,  and it is denoted in these sectors,
\beq
\label{j4.11}
G_{ \mu\nu }^{ A }  =
:  \{ \; \mathcal{ G }_{ \mu\nu }^{ a }  ;  [ Q,  \mathcal{ G }_{ \mu\nu }^{ a } ] \;\! \},
\eeq
where $\mathcal{ T }_{ A }$  is given in (\ref{j4.8}), and
$\mathcal{ G }_{ \mu\nu }^{ a }$ is defined by
\beq
\label{j4.12}
\mathcal{ G }_{ \mu\nu }  =:  \mathcal{ G }^a_{ \mu\nu } \mathcal{ P }_{ a }.
\eeq

\section{The most general case}

Now, let us consider the most general involution (\ref{j17}), without assuming the
conditions (\ref{j18}), (\ref{j19}).  In that case, one should seek for a solution to
the operator $Q$ in the form of a ghost power series expansion of the form
\beq
\label{j22}
Q =  C^{a }t_{a} + \frac{1}{2} C^{b} C^{a} U_{ab}^{c} \mathcal{P}_{c} +
\frac{1}{12} C^{c} C^{b} C^{a} U_{abc}^{ed} \mathcal{P}_{d} \mathcal{P}_{e}  +
\cdots  .  
\eeq We do assume the following irreducibility condition for the
generators $t_{a}$ to be satisfied:
\beq
\label{j23}
Z^{a} \neq 0,
\quad   Z^{a} t_{a} = 0, \quad \Rightarrow\quad
 Z^{a} = Z^{cb} \Pi_{bc}^{a}, 
\eeq where $Z^{cb}  =  - Z^{bc}$  are arbitrary, and we have denoted
\beq \label{j24}
\Pi_{bc}^{a}  =  t_{b} \delta_{c}^{a} -
( b \leftrightarrow c) - U_{bc}^{a},  
\eeq
so that there holds identically
\beq
\label{j25}
\Pi_{bc}^{a} t_{a}  =  [ t_{b}, t_{c} ] - U_{bc}^{a} t_{a}   =  0. 
\eeq
In case the $Z^{a}$, in the first and second in (\ref{j23}), have some extra free
indices, these indices are inherited by the  $Z^{ab}$, in the third in (\ref{j23}),
together with their symmetry properties, if any.
Now, commute the first in (\ref{j17}) with $t_{d}$ and then sum up the cyclic
permutations $(a, b, d)$. By using (\ref{j17}) again, one gets
\beq
\label{j26}
Y_{abd}^{e} t_{e} = 0,  
\eeq
where $Y_{abd}^{e}$ just denotes the left-hand side in (\ref{j18}). Due to the
irreducibility (\ref{j23}), one gets
\beq
\label{j27}
Y_{abd}^{e}  = -\frac{1}{2}\;U_{abd}^{fg} \Pi_{gf}^{e},    
\eeq
which is exactly the relation that does follow from (\ref{j2}) to the  $(CCC
\mathcal{P} \mathcal{P} )$ order. In this way, one is able, in principle, to
show, order by order, that there formally exist all the structure operators in the
series expansion (\ref{j22}).


In the case of a Lie group, where the generators $t_{a}$ and $T_{a}$
do satisfy the same algebra, there exists a natural
counterpart to (\ref{j25}) in terms of $T_{a}$, that is
\beq
\label{j*27}
\tilde{ \Pi }_{ab}^{c}  =  T_{a} \delta_{b}^{c}  -  ( a \leftrightarrow   b )  -
U_{ab} ^{c},  \quad     \tilde{ \Pi }_{ab}^{c}  T_{c}   =   0,    
\eeq
 which extends naturally the irredicibility concept as to the ghost-extended
generators  $T_{a}$. Then, we have the following relation to hold
\beq
\label{j*28}
\tilde{ \Pi }_{ab}^{c}  =  \Pi_{ab}^{c}  +  (  C^{d} U_{da}^{e} \mathcal { P
}_{e} \delta_{b}^{c}  -  ( a  \leftrightarrow  b )   ).     
\eeq

\section{Natural canonical equivalence}

As  to  the  nilpotency condition  (\ref{j2}),  one can always
subject the operator $Q$ to an arbitrary canonical transformation
\cite{BTijmpa}
\beq \label{j6.1}
Q  \;\rightarrow\; Q'  = \exp\{ s G\} Q \exp\{ - s G \}, 
\eeq
where $s$ is a Boson parameter, and $G$ is a matrix valued and ghost dependent
generator,
\beq
\label{j6.2}
\varepsilon ( G )  =  0,   \quad    {\rm gh} ( G )  =  0.       
\eeq
We  have
\beq
\label{j6.3}
&&[ Q' , Q' ]  = 0,\quad  \varepsilon ( Q' )  =  1,  \quad   {\rm gh} ( Q' )   =  1,\\ 
\label{j6.4}
&&\pa_{s} Q'  =  [ G , Q' ] ,   \quad    Q' \big|_{ s = 0 }  = Q,   \quad
\pa_{s}  =  \frac{\pa}{\pa s} ,    \\    
\label{j6.5}
&&Q'=  C^{a} t'_{a} + \frac{1}{2} C^{b} C^{a}\; U_{ab}^{'c}\; \mathcal{P}_{c}  +  ..., \\  
\label{j6.6}
&&G =  G_{0}  +  C^{a}\; G_{a}^{b}\; \mathcal{P}_{b}  +  ...   ,   
\eeq
with all matrix valued  structure coefficients.  It follows from (\ref{j6.3}) that
the all the primed structure  coefficients of the primed $Q'$ satisfy the same
equations as their unprimed counterparts do.  In turn, it follows from (\ref{j6.4})
\beq
\label{j6.7}
\pa_{s} t'_{a}  =  [ G_{0} ,  t'_{a} ]  +  G_{a}^{b} t'_{b},     \quad
t'_{a}\big|_{ s = 0 }  =  t_{a},     
\eeq
\beq
\label{j6.8}
\pa_{s} U_{ab}^{'c}  = [ G_{0} , U_{ab}^{'c} ]  +  ( G_{a}^{d}\; U_{db}^{'c}  -
( a \leftrightarrow b ) )  -  U_{ab}^{'d}\; G_{d}^{c}  +
G_{ab}^{de} \;U_{ed}^{'c},    \quad    U_{ab}^{'c} \big|_{ s = 0 }  =  U_{ab}^{c}, 
\eeq
and so on.  Here in (\ref{j6.8}),  $G_{ab}^{de}$  are structure coefficients as to
the order $CC\mathcal{P}\mathcal{P}$ in  $G$ (\ref{j6.6}).  These equations do
determine the transformation law as to  all the structure coefficients in (\ref{j6.5}).  In
particular, the $G_{0}$ does determine the canonical transformation in the
original  matrix valued sector.  In turn, the  $G_{a}^{b}$ do determine the actual rotations as
to the basis  of the original generators. In turn, the latter  two
transformations,
as induced to  the next structure coefficient $U_{ab}^{c}$,  are  determined
by the equation (\ref{j6.8}), and so on in (\ref{j6.5}).  Our main conjecture claims that
the natural arbitrariness (\ref{j6.1}) is maximal, if the irreducibility (\ref{j23}) holds
for primed basis of the generators $ t'_{a}$, as well. In that case, canonical
transformations (\ref{j6.1}) are capable to interpolate between  the most general
generator and Abelian ones.

If one rewrites the (\ref{j25}) in the form with enumerated indices,
\beq
\label{j6.9}
\Pi_{ a_{1} a_{2} }^{ b_{1} }  t_{ b_{1} }   =   0,     
\eeq
due to the nilpotency (\ref{j2}), it becomes rather obvious that there exists a
chain of recursive relations extending  (\ref{j6.9}) as
\beq
\label{j6.10}
\Pi_{ a_{1} ...  a_{n + 1} }^{ b_{n} ... b_{1} }  \Pi_{ b_{1} ... b _{n} }^{
c_{n - 1} ... c_{1} }   =   0,  \quad   n  =  2, ...  ,         
\eeq
where the n-th  $\Pi$  (with $n $ uppercases ) is constructed of the first $n +1$
structure coefficients in (\ref{j22}).  That chain of recursive relations extends
naturally  the irreducibility concept as to higher structure coefficients.
As an example, we demonstrate the case $n = 2$:
\beq
\label{j6.11}
\Pi _{ a_{1} a_{2} a_{3} }^{ b_{2} b_{1} }   =   \frac{1}{2} (  \Pi_{ a_{1} a _{2}
}^{ b_{2} } \delta_{ a_{3} }^{ b_{1} }  -  ( b_{1}\leftrightarrow b_{2} )  )  +
 {\rm cyclic \; perm}.  ( a_{1}, a_{2}, a_{3} )  -
 \frac{1}{2}\; {\tilde U}_{ a_{1} a_{2} a_{3} }^{ b_{2}b_{1} },                    
\eeq
where we have also used the relation
\beq
\label{j6.12}
\Pi_{ a_{1} a_{2} }^{ b_{2} } \Pi_{ b_{2} a_{3} } ^{ c_{1} }  + {\rm cyclic \;perm}.
 ( a_{1}, a_{2}, a_{3} )  =
- \frac{1}{2}\;
{\tilde U}_{ a_{1} a_{2} a_{3} }^{b_{2} b_{1}}  \Pi_{ b_{1} b_{2} }^{ c_{1} },   
\eeq
\beq
\label{j6.12}
\frac{1}{2}(  U  -  \tilde{U}  )_{ a_{1} a_{2} a_{3} }^{ b_{2} b_{1} }
\Pi_{ b_{1}b_{2} }^{ c_{1} }   =
(  U_{ a_{1} a_{2} } ^{ b_{1} }  t_{ b_{1} } \delta_{a_{3} }^{ c_{1} }  +
t_{ a_{2} } U_{ a_{1} a_{3} }^{ c_{1} }  )  +  {\rm cyclic \; perm}. ( a_{1},
a_{2}, a_{3} ).    
\eeq

One can resolve for the $\tilde{U}$  operators,
\beq
\label{j6.14}
\frac{1}{2} ( U - \tilde{U} )_{a_{1} a_{2} a_{3} }^{ b_{2} b_{1} }  =
\left[\frac{1}{2} (
\delta_{ a_{1} }^{ b_{2} } t_{ a_{2} } \delta_{ a_{3} }^{ b_{1} }  -  (
b_{1}\leftrightarrow b_{2} )  )  +
{\rm cyclic\; perm}. ( a_{1}, a_{2}, a_{3} ) \right],       
\eeq
to get the following explicit solution
\beq
\nonumber
\label{j6.15}
&&\Pi_{ a_{1} a_{2} a_{3} }^{ b_{2} b_{1} }  =  \left[  t_{ a_{1} } \frac{1}{2} (
\delta_{ a_{2} }^{ b_{2} } \delta_{ a_{3} }^{ b_{1} }  -  ( b_{1}\leftrightarrow
b_{2} ) ) +
{\rm cyclic\; perm}. ( a_{1}, a_{2}, a_{3} )  \right]  - \\
&&\qquad\qquad\!\!\! - \left[\frac{1}{2} ( U_{ a_{1} a_{2} }^{
b_{2} } \delta_{ a_{3} }^{ b_{1} }  -  ( b_{1}\leftrightarrow b_{2} ) ) +
{\rm cyclic\; perm}. ( a_{1}, a_{2}, a_{3} )  \right]  -
\frac{1}{2}\;U_{ a_{1} a_{2} a_{3} }^{b_{2} b_{1} }.     
\eeq
\\

\section{Note added in proof}

Here we claim that the standard Faddeev-Popov measure can also be naturally
reformulated in terms of the generators $T_{a}$ (\ref{j2.24}), by using the representation
similar to (\ref{j1}) as applied to the Nakanishi-Lautrup matrix valued fields
$\Pi$ ( Lagrange multipliers for
gauge fixing functions),
as well  as to the ghost and antighost Faddeev-Popov matrix valued field
$B, \bar{B}$. Then, in the
case of the Lorentz gauge, $\pa_{\mu} A_{\mu}  =  0$,  the  gauge fixing part
of the total Lagrangian reads
\beq
\label{N.1}
 (N n)^{-1}  \tr ( \Pi\! \;\pa_{\mu} A_{\mu}  +
( \pa_{\mu} \bar{B} )  ( \pa_{\mu} B  +  [ A_{\mu},  B ] ) ), 
\eeq
where all fields take their values in the $T$- algebra,
\beq
\label{N.2}
A_{\mu}  =  A_{\mu}^{a} T_{a}, \quad \Pi  =  \Pi^{a} T_{a}, \quad  B  =  B^{a} T_{a},
\quad \bar{B}  =  \bar{B}^{a} T_a .    
\eeq
The (\ref{N.1}) is invariant under the standard BRST transformations
\beq
\label{N.3}
\delta A_{\mu}  =  ( \pa_{\mu} B  +  [ A_{\mu}, B ] ) \mu,\quad
\delta B  =  \frac{1}{2} [ B, B ] \mu, \quad  \delta \bar{B}  =  - \Pi \mu,
\quad \delta\Pi  =  0.    
\eeq
If the $T_{a}$ do satisfy a Lie algebra, the BRST invariance holds in  a
straightforward way, with all the
coefficients in (\ref{N.2}) (field components) being $c$- numbers. However, in the
quasigroup /groupoid case, one should allow for these coefficients
to be matrix valued, in general. Then, we have from
(\ref{j1}) and the first in  (\ref{j2.24}) and (\ref{j2.27})
\beq
\label{N.4}
A_{\mu}  =   A_{\mu}^{a} T_{a} + [ Q, A_{\mu}^{a} ] \mathcal{P}_{a}, 
\eeq
and similar formulae for all other fields. The form of the second term here
is quite similar to the one of the second term in (\ref{j4.6}),
that makes unclosed the algebra of the generators $T_{a}$ alone.
The doubled generators $\mathcal{T}_{A}$, (\ref{j4.8}), do satisfy the closed
involution
\beq
\label{N.5}
[ \mathcal{T}_{A}, \mathcal{T}_{B} ]  =  \mathcal{U}_{AB}^{\;C}
\mathcal{T}_{C},      
\eeq
with the structure coefficients $\mathcal{U}_{AB}^{\;C}$  given explicitly in
the  relations (\ref{j4.6}), (\ref{j4.7}).

Any  operator $\mathcal{X}$ of the form similar to (\ref{N.4}),
\beq
\label{N.6}
\mathcal{X}  =  \mathcal{X}^{A} \mathcal{T}_{A},\quad
\mathcal{X}^{A}  =  \{ X^{a} (-1)^{ \varepsilon_{ \mathcal{X} } } ; [ Q,
X^{a} ] \},          
\eeq
with *matrix valued* coefficients $\mathcal{X}^{A}$, belongs to the closed
doubled $\mathcal{T}$- algebra. The latter makes all the  commutators
 entering  (\ref{N.1}), (\ref{N.3})
well defined as taking their values within the same  closed
doubled $\mathcal{T}$- algebra.  Notice
that the (\ref{N.6}) rewrites in the natural form
\beq
\label{N.7}
\mathcal{X}  =  [ Q, X^{a} \mathcal{P}_{a} ],   
\eeq
maintained under commuting  of two operators of the form  (\ref{N.7}), due to the
ghost number conservation.

\section*{Acknowledgments}
\noindent
 The work of I. A. Batalin is
supported in part by the RFBR grants 14-01-00489 and 14-02-01171.
The work of P. M. Lavrov is supported in part by  the Presidential grant
88.2014.2 for LRSS and the FRBR grant 15-02-03594.
\\

\appendix
\section*{Appendix A. Generating equations for the quantum antibracket\\ algebra}
\setcounter{section}{1}
\renewcommand{\theequation}{\thesection.\arabic{equation}}
\setcounter{equation}{0}

Here we include in short the generating equations for the quantum
antibracket algebra \cite{BM1}.  Let us introduce an operator valued exponential
\beq
\label{A1}
U = \exp\{ \lambda^{a} f_{a} \},    \quad   U|_{\lambda = 0} = 1,       
\eeq
 where $\{ f_{a},  a = 1, 2, ...  \}$,  is a chain of operators,  $\varepsilon(
f_{a} ) = \varepsilon_{a}$,  and $\lambda^{a}$ are parameters, $\varepsilon(
\lambda^{a} )  =  \varepsilon_{a}$.  Introduce the
$U$-transformed $Q$-operator,
\beq
\label{A2}
\tilde{Q}  =  U Q U^{-1},   \quad  \tilde{Q}^{2} = 0.       
\eeq
 We have
\beq
\label{A3}
\pa_{a} \tilde{Q}  =  [ R_{a}, \tilde{Q} ],   \quad   R_{a} =  ( \pa_{a} U )
U^{-1},    \quad     \pa_{a} = \frac{\pa}{\pa\lambda^{a}}, \quad
     \tilde{Q}|_{\lambda = 0} = Q,   
\eeq
 \beq
\label{A4}
\pa_{a} R_{b}  -  \pa_{b} R_{a} (-1)^{ \varepsilon_{a} \varepsilon_{b} }  =  [
R_{a}, R_{b}] .    
\eeq

The Lie equation (\ref{A3}) and the  Maurer-Cartan equation
(\ref{A4}) do serve as the generating equations for quantum
antibrackets.  Here we present explicitly only the case of quantum
2-antibracket.   It follows from (\ref{A3}) by  $\lambda$
differentiating , that \beq \nonumber &&- \pa_{a} \pa_{b} \tilde{Q}
(-1)^{\varepsilon_{b}}  + \frac{1}{2} [ ( \pa_{a} R_{b} + \pa_{b}
R_{a} (-1)^{ \varepsilon_{a}
\varepsilon_{b} } ) (-1)^{\varepsilon_{b}} ,  \tilde{Q}  ]  =  \\
\label{A5}
&&=\frac{1}{2} \left(  [ R_{a}, [ \tilde{Q}, R_{b} ] ]  -  ( a \leftrightarrow b ) (-1)^{ (
\varepsilon_{a} + 1 ) ( \varepsilon_{b} + 1 ) }  \right) =  ( R_{a}, R_{b} ) _{
\tilde{Q} }.      
\eeq
 It follows from (\ref{A5}) at $\lambda = 0$,
\beq
\label{A6}
-  ( \pa_{a}\pa_{b} \tilde {Q} ) (-1)^{ \varepsilon_{b} } |_{\lambda = 0}  =
( f_{a}, f_{b} )_{Q},     
\eeq
 where we have used
\beq
\label{A7}
 ( \pa_{a} R_{b} ) |_{\lambda = 0}  =  \frac{1}{2}  [ f_{a}, f_{b} ].   
\eeq
 It follows in a similar way that higher $\lambda$  derivatives of $\tilde{Q}$ do
yield all higher quantum antibrackets,
\beq
 ( f_{ a_{1} } ,  ...  , f_{ a_{n} } )_{Q}   =  - {\rm Sym}( [ f_{
a_{1} }, ...  , [ f_{ a_{n} }, Q ] ... ] )  (-1)^{ E_{n} }, 
\eeq
where  we have denoted
\beq
E_{n} =  \sum_{ k = 1 }^{ [ n/2 ] } \varepsilon_{ a_{2k} }, 
\eeq
\beq
{\rm Sym}( X_{ a_{1}  ...  a_{n} } )  =  S_{ a_{1}  ...  a_{n} }^{ b_{n}
... b_{1} }  X_{ b_{1}  ...  b_{n} },    
\quad n!  S_{ a_{1}  ...  a_{n} }^{ b_{n}  ...  b_{1} }  =  \pa_{ a_{1} }
... \pa_{ a_{n} }  \lambda^{ b_{n} }  ...  \lambda^{ b_{1} }.
\eeq
It has also been shown in \cite{BM1}, how these equations enable one
to derive the modified Jacobi relations for subsequent higher
quantum antibrackets.

Notice, in conclusion, that there exists a nice interpretation of
the quantum antibracket algebra via the so-called differential
polarization \cite{Ber}. In particular, being $B$ an arbitrary Boson
operator, one can 3-times  commute that $B$ with the nilpotency
equation (\ref{j2}), to get the relation
\beq
\label{A11}
6 ( B, ( B, B )_{Q})_{Q}  =  [ ( B, B, B )_{Q}, Q ] , \quad \varepsilon( B )  = 0.
\eeq
Then, by choosing $B$ in the form
\beq
\label{A12}
B  =  \alpha X + \beta Y + \gamma Z,   
\eeq
with parameters $\alpha, \beta, \gamma$   of the same Grassmann parities as the ones
of the operators $X, Y, Z$, respectively, are, one applies to (\ref{A11})
the differential  operator
\beq
\label{A13}
\pa_{\alpha} \pa_{\beta} \pa_{\gamma} (-1)^{ ( \varepsilon_{\alpha} + 1) ( \varepsilon_{\gamma}
+ 1 ) + \varepsilon_{\beta} },     
\eeq
to get exactly the relation (\ref{j15*}).


\begin {thebibliography}{99}
\addtolength{\itemsep}{-8pt}

\bibitem{YM}
C. N. Yang,  R. L. Mills, Conservation of isotopic spin and isotopic gauge
invariance,  Phys. Rev. 96 (1954) 191.

\bibitem{DeW1}
B. S. De Witt, Quantum theory of gravity. I.
The canonical theory,
Phys. Rev. 160 (1967) 1113.


\bibitem{DeW}
B. S. De Witt, Quantum theory of gravity. II.
The manifestly covariant theory,
Phys. Rev. 162 (1967) 1195.

\bibitem{FP}
L. D. Faddeev, V. N. Popov, Feynman diagrams for the Yang-Mills field,
Phys. Lett. B 25 (1967) 29.

\bibitem{FT}
E. S. Fradkin, I. V. Tyutin, S matrix for Yang-Mills and gravitational fields,
Phys. Rev. D 2 (1970) 2841.

\bibitem{T}
J. C. Taylor, Ward identities and charge renormalization
of the Yang-Mills field,
Nucl. Phys. B 33 (1971) 436.

\bibitem{S}
A. A. Slavnov, Ward identities in gauge theories,
Theor. Math. Phys. 10 (1972) 99.

\bibitem{DeWitt}
B. S. De Witt, Dynamical theory of groups and fields, Gordon and Breach, 1965.

\bibitem{brs1}
C. Becchi, A. Rouet, R. Stora, The abelian Higgs Kibble Model,
unitarity of the $S$-operator,
Phys. Lett. B 52 (1974) 344.

\bibitem{t}
I. V. Tyutin, Gauge invariance in field theory and statistical physics
in operator formalism, Lebedev Institute preprint  No.  39  (1975),
arXiv:0812.0580 [hep-th].

\bibitem{KO}
T. Kugo, I. Ojima, Local covariant operator formalism
of non-abelian gauge theories and quark confinement problem,
Progr. Theor. Phys. Suppl. 66 (1979) 1.

\bibitem{BV}
I. A. Batalin , G. A. Vilkovisky, Gauge algebra and quantization,
Phys. Lett. B 102 (1981) 27.

\bibitem{BF1}
I. A. Batalin, E. S. Fradkin, Operator quantization of relativistic
dynamical systems subject to first class constraints,
Phys. Lett. B 128 (1983) 303.

\bibitem{BF2}
I. A. Batalin, E. S. Fradkin, Operator quantization of
dynamical systems with  first class and second class constraints,
Phys. Lett. B 180 (1986) 157.

\bibitem{Sab}
L. V. Sabinin, Methods of non-associative algebra in differential geometry,
{\it in:} S. Kobayashi, K. Nomizu, {\it Foundations of differential geometry},
Interscience  Publishing, New-York - London, v.1, 1963.


\bibitem{M}
A. I. Mal'cev, Algebraic systems, Moscow, Nauka, 1970 (in Russian).

\bibitem{B}
I. A. Batalin, Quasigroup construction and first class constraints,  J.
Math. Phys. 22  (1981) 1837.

\bibitem{BVjmp}
I. A. Batalin , G. A. Vilkovisky,
Existence theorem for gauge algebra,
J. Math. Phys. 26 (1985) 172.

\bibitem{CFer}
M. Crainic, R. L. Fernandes, Integrability of Lie brackets,
Ann. of Math. 157 (2003) 575. 

\bibitem{Ber}
K. Bering,  Non-commutative Batalin -Vilkovisky  algebras, strongly
homotopy Lie algebras, and the Courant bracket,
Comm. Math. Phys. 274 (2007) 297.

\bibitem{MSt}
C. Mayer, T. Strobl, Lie Algebroid Yang Mills with matter fields,
 J. Geom. Phys. 59 (2009) 1613.

\bibitem{GSt}
M. Grutzmann, T. Strobl, General Yang-Mills type gauge theories
for p-form gauge fields: From physics-based ideas to a mathematical
framework or From Bianchi identities to twisted Courant algebroids,
Int. J. Geom. Meth. Mod. Phys. 12 (2014) 01, 1550009.

\bibitem{BM}
I. Batalin, R. Marnelius, Quantum antibrackets, Phys. Lett. B 434
(1998) 312.

\bibitem{BM1}
I. Batalin, R. Marnelius, General quantum antibrackets,
Theor. Math. Phys.  120 (1999) 1115.

\bibitem{K-S}
Yv. Kosmann-Schwarzbach, Derived brackets,
Lett. Math. Phys. 69 (2004) 61.

\bibitem{Vo}
Th. Voronov, Higher derived brackets and homotopy algebras,
J. Pure Appl. Algebra 202 (2005) 133.

\bibitem{CSch}
A. S. Cattaneo, F. Schatz, Equivalence of higher derived brackets,
J. Pure Appl. Algebra 212  (2008) 2450.


\bibitem{Hw}
S. Hwang, Covariant quantization of the string in dimensions D
$\leq$ 26 using a BRS formulation, Phys. Rev. D 28 (1983)
2614.

\bibitem{GShW}
M. B. Green,  J. H. Schwarz, E.  Witten, Superstring Theory,
vol. 1, Cambridge University Press, 1987.


\bibitem{BTtmf}
I. A. Batalin,  I. V. Tyutin,  BRST - invariant algebra  of
constraints in terms of commutators and quantum antibrackets,
Theor. Math. Phys. 138 (2004) 1.


\bibitem{BB}
I. A. Batalin,  K.  Bering, Reducible gauge algebra of  BRST -
invariant constraints,  Nucl. Phys. B 711 [PM] (2007) 190.


\bibitem{BTijmpa}
I. A. Batalin,  I. V. Tyutin, On the transformations of
Hamiltonian gauge algebra under rotations of constraints, Int. J.
Mod. Phys. A 20 (2005) 895.



\end{thebibliography}
\end{document}